\documentclass{appolb}
\usepackage[T1]{fontenc}
\usepackage[latin9]{inputenc}
\usepackage{amsmath}
\usepackage{graphicx}
\usepackage{esint}

\makeatletter

\providecommand{\tabularnewline}{\\}

\newcommand{\diracsl}{\displaystyle{\not} }
\usepackage{psfrag}

\makeatother

\begin{document}

\title{Low-energy amplitudes in the non-local chiral quark model}
\author{
Piotr Kotko{\thanks{e-mail: kotko@th.if.uj.edu.pl}
\thanks{Presented at the 49-th Cracow School of Theoretical Physics, 
May 31 -June 10, 2009, Zakopane, Poland}}
\\
\emph{M.Smoluchowski Institute of Physics,} \\
\emph{Jagiellonian University,} \\
\emph{Reymonta 4, 30-059 Krak\'{o}w, Poland.}}
\date{\today}

\maketitle
\begin{abstract}
We apply chiral quark model with momentum dependent quark mass to two kinds of 
nonperturbative objects. These are:  photon Distribution Amplitudes which we calculate
up to twist-4 in tensor, vector and axial channels and pion-photon Transition Distribution Amplitudes together with related form factors. 
Where possible we compare our results with experimental data.
\end{abstract}

\section{Introduction}

\label{sec:introduction}

One of the biggest problems in particle physics is description of
hadrons in terms of the fundamental degrees of freedom --- quarks and
gluons. This is in fact a non-perturbative problem and usually is
formulated in terms of various distribution functions, which appear
in QCD factorization theorems. The most famous example are Parton
Distribution Functions which can be measured in the inclusive lepton-hadron
deep inelastic processes. They are, however, one dimensional distributions
only. Therefore although they are by now sufficient for description
of various processes at high energies, they simply
give only limited information on the structure of hadrons. 

On the other hand one can study also hard \textit{exclusive processes},
such as deeply virtual Compton scattering for instance. Then the factorization
theorem states, in great simplicity, that the amplitude is given by
the convolution
\begin{equation}
\mathcal{M}=\left(soft\right)\otimes\left(hard\right),\label{eq:factoriz}
\end{equation}
where $hard$ is the part that can be calculated in perturbative QCD,
while the $soft$ part is of non-perturbative nature. In the following
we will be mainly concentrated on the $soft$ part. Although
difficult to access experimentally, they can be obtained either by lattice calculations or --- as we shall see
--- they can be estimated from theoretically justified low energy effective
models. 

The $soft$ part parametrizes hadronic matrix elements of certain
non-local operators on the light-cone. The simplest objects of this
kind are Distribution Amplitudes (DA) which correspond to hadron-to-vacuum
matrix elements of bi-local quark operators on the light-cone. In case
of the leading twist DAs, \textit{i.e.} the ones giving main contribution
to the amplitude, they describe (in the infinite momentum frame) the
probability for a composite particle to dissociate into
its constituents with given longitudinal momentum fractions. Distribution
Amplitudes have been successfully used in theoretical description of
hadronic form factors \cite{Radushkin,Brodsky} for many years. However
recent BaBar data for pion-photon transition form factor shows, that probably 
the standard factorization formulae do not apply \cite{RadyushkinBABAR,PolyakovBABAR}.
We shall come back to the BaBar data in Section \ref{sec:TDA}.

More general class of $soft$ objects are Generalized Parton Distributions
(GPD). They correspond to non-diagonal in momenta matrix elements,
therefore they describe also the distributions of transverse momenta
of the partons inside the hadron. GPDs appear in description of deeply
virtual Compton scattering for instance, which is recently
the subject of intensive theoretical and experimental studies. For a review
of this issue see \textit{e.g.} \cite{Belitsky}.

One can still define more general class of the objects than GPDs
--- so called Transition Distribution Amplitudes (TDA). They parametrize
matrix elements which are non-diagonal in momenta and in physical states.
Such a family of objects was introduced for the first time in \cite{Pire}. We shall discuss this class further in Section \ref{sec:TDA}.

As already remarked above there is very little experimental data concerning
the $soft$ part of \eqref{eq:factoriz}. On the other hand it can be studied in
effective models. This is however nontrivial not only because of complex
non-local interactions at low energies. Even bigger problem is that
in general effective models do not inherit all symmetries of the underlying
theory. Soft objects considered here appear in the framework of QCD, therefore
they should posses several important properties, for example Lorentz
and gauge invariance. They should also correctly reproduce quantum anomalies.
The task to cope with all the constrains in the effective models is therefore
nontrivial. We shall come back to this point in Section \ref{sec:model}.

\section{Non-local chiral quark model}

\label{sec:model}

Let us consider the scattering process involving the simplest possible
hadronic state --- the pion. On one hand it is a bound state of
quark-anti-quark pair and the Goldstone boson of spontaneously broken
chiral symmetry on the other. Before we proceed let us briefly recall
these very important aspects of QCD. 

Here and in the following we assume only two quarks $u$ and $d$
which are massless, \textit{i.e.} $m_{u}=m_{d}=0$. Then the Lagrangian
of QCD is invariant under separate rotations of left- and right-handed
spinors, that is the symmetry group is $\mathrm{SU}\left(2\right)_{\mathrm{R}}\otimes 
\mathrm{SU}\left(2\right)_{\mathrm{L}}$
(the chiral symmetry group). It is generated by the chiral charges
satisfying $\mathrm{SU}\left(2\right)$ commutation relations $Q_{\mathrm{L},\mathrm{R}}^{a}=\int d^{3}x\,\psi_{\mathrm{L},\mathrm{R}}^{\dagger}\left(x\right)\gamma_{5}\frac{\tau^{a}}{2}\psi_{\mathrm{L},\mathrm{R}}\left(x\right)$,
where $\tau^{a}$ are Pauli matrices, $\psi$ denote iso-doublets.
One can also define the combination of chiral fields transforming
as vector and axial-vector. Then the corresponding combination of
$\mathrm{L},\mathrm{R}$ charges $Q^{a}=Q_{\mathrm{R}}^{a}+Q_{\mathrm{L}}^{a}$ and $Q_{5}^{a}=Q_{\mathrm{R}}^{a}-Q_{\mathrm{L}}^{a}$
generate the $\mathrm{SU}\left(2\right)_{\mathrm{V}}\otimes \mathrm{SU}\left(2\right)_{\mathrm{A}}$ group%
\footnote{There is also similar global symmetry acting on the whole doublet.
The axial symmetry $\mathrm{U}\left(1\right)_{\mathrm{A}}$ is however broken due to
quantum anomaly.%
}. The most direct consequence of this symmetry would be a degeneracy
of the states with different parity. However, such a behavior is not
seen in the hadronic spectrum --- on the contrary, we observe huge mass
differences between parity partners. The most natural way of solving
this discrepancy, is to postulate that although theory is chirally
invariant, the vacuum state is not. This phenomenon is known as \textit{spontaneous}
\textit{chiral symmetry breaking} (S$\chi$SB). According to Goldstone
theorem we should then observe a triplet of massless pseudo-scalar
particles --- the Goldstone bosons. Indeed they can be apparently identified
with pions ($\pi^{+}$, $\pi^{0}$, $\pi^{-}$), which are very light
($m_{\pi}\approx140\,\mathrm{MeV}$) in comparison to other hadrons
(e.g. $m_{\mathrm{proton}}\approx1\,\mathrm{GeV}$). Non-zero pion
mass can be explained by finite (although small) current masses of $u,\, d$
quarks, which explicitly break chiral symmetry from the very beginning. 

Another important aspect of S$\chi$SB is the existence of the quark condensates,
\textit{i.e.} the quantities \begin{equation}
\left\langle 0\left|\bar{q}q\right|0\right\rangle \equiv\left\langle \bar{q}q\right\rangle =\left\langle \bar{q}_{R}q_{L}\right\rangle +\left\langle \bar{q}_{L}q_{R}\right\rangle ,\end{equation}
where $q$ denotes either $u$ or $d$ quark field. It can be easily
seen that the nonzero value of the quark condensate breaks chiral
symmetry of the vacuum. Consider the commutator\begin{equation}
\left[Q_{5}^{a},\bar{\psi}\gamma_{5}\tau^{b}\psi\right]=-\delta^{ab}\bar{\psi}\psi\end{equation}
and its vacuum expectation value. If the right hand side is nonzero
it implies that\begin{equation}
Q_{5}^{a}\left|0\right\rangle \neq0,\end{equation}
what is exactly the S$\chi$SB condition. Therefore the quark condensate
can be viewed as an order parameter measuring the breakdown of chiral
symmetry. Phenomenological value of the quark condensate is quite
large $\left\langle \bar{q}q\right\rangle \sim\left(-250\,\mathrm{MeV}\right)^{3}$
(at renormalization scale about $1\,\mathrm{GeV}$). Let us notice
next that in QCD $\left\langle \bar{q}q\right\rangle $ is represented
by a closed quark loop, \textit{i.e.} it is proportional to the trace of fermionic propagator
$\left\langle \bar{q}q\right\rangle \sim\mathrm{Tr}\,\hat{S}\left(x,x\right)$,
where the trace is over Dirac and color indices. However, if this quantity
is non-zero there must be a non-slash term in the propagator --- the
mass term. This dynamically (due to S$\chi$SB) generated mass is
often referred to as \textit{constituent quark mass}. Notice that
the quark condensate is a purely non-perturbative quantity, since
it is impossible to generate non-slash quark self energy by interactions
of vector bosons. Rather it must be created by some kind of a scalar
interactions. We shall come back to the issue of quark condensate
later in Section \ref{sec:foton}.

Let us now switch to description of the interactions between the pions
and quarks at low energies. It is clear from the above that such a model
must incorporate S$\chi$SB. It is convenient to start discussion
by recalling the famous Nambu---Jona-Lasinio (NJL) model. It is an effective
theory of quarks with four fermion couplings, appearing due to integrating
out the gluonic degrees of freedom from the QCD action. In the standard NJL model couplings
with more fermions are neglected. The Lagrange density for the simplest
version of the NJL model reads\begin{equation}
\mathcal{L}_{\mathrm{NJL}}=\bar{\psi}i\diracsl\partial\psi+\frac{G}{2}\left[\left(\bar{\psi}\psi\right)^{2}+\left(\bar{\psi}i\gamma_{5}\vec{\tau}\psi\right)^{2}\right],\label{eq:NJL}\end{equation}
where $\vec{\tau}=\left(\tau^{1},\tau^{2},\tau^{3}\right)$ are Pauli
matrices and $G$ is coupling constant. It can be checked using
some algebra and the relation $e^{-i\left(\alpha\cdot\tau\right)\gamma_{5}}=\left(\cos\left|\alpha\right|-i\gamma_{5}\hat{\alpha}\cdot\tau\sin\left|\alpha\right|\right),$
where $\alpha^{i}=\left|\alpha\right|\hat{\alpha}^{i}$ that Lagrangian
\eqref{eq:NJL} is indeed chirally invariant. The most important feature
of NJL model is that it incorporates the mechanism leading to S$\chi$SB.
One way to see this is to solve the corresponding lowest order Dyson-Schwinger
equation for the quark propagator. Denoting quark self-energy by $\Sigma\left(p\right)\equiv M$
one obtains the following consistency condition (so called gap equation)\begin{equation}
M=-i8GN_{c}\int\frac{d^{4}k}{\left(2\pi\right)^{4}}\frac{1}{k^{2}-M^{2}}.\label{eq:gap}\end{equation}
It has two solutions: $M=0$ (for massless quarks) and $M\neq0$.
The latter corresponds to the constituent quark mass which generates
non-zero quark condensate breaking the chiral symmetry of the vacuum.
Notice that the integral in \eqref{eq:gap} requires regularization. We
shall discuss this later in this section. For a review of NJL model
see \cite{Klevansky} for example.

Mesons can be easily introduced into the just described theory
as auxiliary fields $\sigma$ and $\pi^{a}$ --- this can be
done formally in the path integral formalism and is called bosonization
procedure. The new Lagrange density reads\begin{equation}
\mathcal{L}_{\mathrm{NJL}'}=\bar{\psi}i\diracsl\partial\psi+g\,\bar{\psi}\left[\sigma+i\gamma_{5}\vec{\tau}\cdot\vec{\pi}\right]\psi+\frac{\mu^{2}}{2}\left(\sigma^{2}+\vec{\pi}^{2}\right),\label{eq:semi_NJL}\end{equation}
where $g^{2}=\mu^{2}G$. Notice that the fields $\sigma,\pi^{a}$
are truly auxiliary - there are no corresponding kinetic terms, moreover
they are composed fields what can be immediately seen using equations
of motion. NJL Lagrangian in the form \eqref{eq:semi_NJL} can also
be used to show that the ground state which minimizes the energy is
populated by the scalar quark condensate.

One can also look at the appearance of mesonic fields from a slightly
different point of view. Consider the following effective Lagrange
density \begin{equation}
\mathcal{L}=\bar{\psi}\left(i\diracsl\partial-M\right)\psi,\label{eq:model_1}\end{equation}
which leads to Dirac equation for the quark with constituent
quark mass $M$. However \eqref{eq:model_1} is obviously not chirally
invariant. In order to fix this deficiency one has to introduce additional fields
in the form\begin{equation}
U^{\gamma_{5}}\left(x\right)=e^{\frac{i}{F_{\pi}}\vec{\tau}\cdot\vec{\pi}\left(x\right)\gamma_{5}}\approx1+\frac{i}{F_{\pi}}\gamma_{5}\tau^{a}\pi^{a}\left(x\right)+\ldots\label{eq:u_field}\end{equation}
where $F_{\pi}\approx93\,\mathrm{MeV}$ is the pion weak decay constant,
and couple them to quarks,\begin{equation}
\mathcal{L}=\bar{\psi}\left(i\diracsl\partial-MU^{\gamma_{5}}\right)\psi.\label{eq:model_2}\end{equation}
Then the axial transformations of quark fields can be absorbed by
pion fields $\pi^{a}$. The Lagrange density
\eqref{eq:model_2} is a starting point for our further considerations
and represents the simplest \textit{local chiral quark model}. It
describes quarks having dynamically generated constituent mass $M$
and interacting with the external pion fields.

The effective theories just described are a non-renormalizable ones.
The regularization introduced in order to make the loop integrals
finite cannot be removed at the very end of the calculations and the
observables depend on its actual form. Moreover it is somehow (but
not straightforwardly) related to the domain of applicability of the
model. There are many ways of regularizing the loop integrals. One
could use for example simple four-momentum cutoff or Pauli-Villars
regularization. However, the point is that the regularization scheme should respect
all symmetries of the underlying theory, \textit{i.e.} QCD. This is extremely
important especially in the case of soft matrix elements as stated
in Section \ref{sec:introduction}. Therefore four-momentum cutoff
is excluded in the first place since it violates Lorentz invariance.
Also very often used Pauli-Villars regularization is not the best
method, because in order to get the results consistent with QCD one
must keep it finite in some diagrams and remove in others (connected
with anomalous processes). On the other hand notice that in reality
the constituent quark mass should not be a constant --- it should vanish
for large quark momenta due to asymptotic freedom. Therefore in
the following we assume that \begin{equation}
M\equiv M\left(k\right)=M_{0}F^{2}\left(k\right),\,\textrm{where }F\left(k\right)\underset{k\rightarrow\infty}{\longrightarrow}0, F\left(0\right)=1.\label{eq:const_mass}\end{equation}
The constituent quark mass at zero momenta $M_{0}$ is chosen to be
about $M_{0}\sim350\,\mathrm{MeV}$. 

The interaction part of the effective action corresponding to \eqref{eq:model_2}
with assumption \eqref{eq:const_mass} can be written in momentum
space as\begin{equation}
S_{\mathrm{int}}=M_{0}\int\frac{d^{4}k\, d^{4}l}{(2\pi)^{8}}\bar{\psi}(k)F\left(k\right)U^{\gamma_{5}}(k-l)F\left(l\right)\psi(l).\label{eq:eff_action}\end{equation}
The explicit shape of $F\left(k\right)$ cannot be obtained from the
gap equation itself. However the action \eqref{eq:eff_action} was
actually obtained in the instanton model of the QCD vacuum, together
with the expression for $F\left(k\right)=F_{\mathrm{inst}}\left(k\right)$.
Unfortunately $F_{\mathrm{inst}}\left(k\right)$ turns out to be a
highly non-trivial function of Euclidean momenta \cite{Diakonov}.
Therefore instead of $F_{\mathrm{inst}}\left(k\right)$ we shall use
the following simple formula in Minkowski space \cite{Rostw}
\begin{equation}
F(k)=\left(\frac{-\Lambda_{n}^{2}}{k^{2}-\Lambda_{n}^{2}+i\epsilon}\right)^{n},\label{Fkdef}
\end{equation}
which reproduces $F_{\mathrm{inst}}$ quite well when continued to
Euclidean space. The parameter $n$ is responsible for the actual
shape of $F\left(k\right)$, therefore we can investigate the sensitivity
of calculated quantities to the form of the cutoff function. The cutoff parameter $\Lambda_{n}$ is adjusted
in such a way that pion decay constant $F_{\pi}$ given by the formula
\cite{Birse}\begin{equation}
F_{\pi}^{2}=\frac{N_{c}}{4\pi^{2}}\int_{0}^{\infty}dk_{\mathrm{E}}^{2}\, k_{\mathrm{E}}^{2}\, \frac{M^{2}\left(k_{\mathrm{E}}\right)-k_{\mathrm{E}}^{2}M\left(k_{\mathrm{E}}\right)M^{\prime}\left(k_{\mathrm{E}}\right)+k_{\mathrm{E}}^{4}M^{\prime}\left(k_{\mathrm{E}}\right)^{2}}{\left(k_{\mathrm{E}}^{2}+M^{2}\left(k_{\mathrm{E}}\right)\right)^{2}}\label{eq:Fpi}\end{equation}
is equal to the experimental value. In the above equation $k_{\mathrm{E}}$
corresponds to Euclidean momentum, while the prime to differentiation
with respect to $k_{\mathrm{E}}^{2}$. For example for constituent
quark mass $M_{0}=350\,\mathrm{MeV}$ and $n=1$ \eqref{eq:Fpi} gives $\Lambda_{1}=836\,\mathrm{MeV}$.

Although momentum dependent quark mass seems to be the most natural
regulator it introduces a serious difficulty. Namely Ward-Takahashi
identities are not satisfied in such a model. It can be most easily
seen by considering a divergence (in momentum space) of the vector
current and applying Dirac equation. This violation turns out to be
not very large, it can however spoil some important properties of
soft matrix elements, like correct normalization for example. In order
to fix this problem the standard vector $\gamma^{\mu}$ and axial
$\gamma^{\mu}\gamma_{5}$ vertices have to be modified by adding new
non-local terms. The problem is, however, that such a modification is
not unique \cite{Birse} (Ward identities fix only longitudinal part
of the vertices). In this work we use the following modified vector
and axial vertices:
 \begin{equation}
\Gamma^{\mu}\left(k,p\right)=\gamma^{\mu}-\frac{k^{\mu}+p^{\mu}}{k^{2}-p^{2}}\left(M\left(k\right)-M\left(p\right)\right),\label{eq:vector_vertex}\end{equation}
\begin{equation}
\Gamma_{5}^{\mu}\left(k,p\right)=\gamma^{\mu}\gamma_{5}+\frac{p^{\mu}-k^{\mu}}{\left(p-k\right)^{2}}\left(M\left(k\right)+M\left(p\right)\right)\gamma_{5},\label{eq:axial_vertex}\end{equation}
reproducing Word-Takahashi identities. The vector vertex does not
introduce additional singularities, while the axial one has a pole
corresponding to the massless pion as it should be \cite{Holdom}. Let us remark
that eq. \eqref{eq:Fpi} expressing $F_{\pi}$ is determined unambigously
since it involves only the derivative of the axial current.

At the end of this section we remark that any non-local (\textit{i.e.} with
momentum dependent constituent quark mass $M\left(k\right)$) chiral
quark model is determined by specifying both the $M\left(k\right)$
and the precise form of all vertices.

\section{Photon Distribution Amplitudes}

\label{sec:foton}

As already remarked in the Introduction the simplest $soft$ objects
are Distribution Amplitudes. In this section we present how the non-local
chiral quark model can be applied to this class. However, instead
of considering the hadronic DA we shall discuss less known photon
DA. This is possible due to the fact that --- besides standard perturbative
part --- photons possess also hadronic component. This fact is very
well known from photoproduction processes, where photon structure
function has to be taken into account (so called resolved photoproduction).

\begin{figure}
\begin{centering}
\label{fig:DA1}\begin{tabular}{ll}
a)  & b)\tabularnewline
 \psfragscanon  \psfrag{DA}{DA} \psfrag{1}{$D^{0\,*}$} \psfrag{2}{$D^{0}$} \psfrag{3}{$\gamma(P)$}\includegraphics[width=3.5cm]{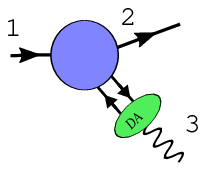}\psfragscanoff
$\qquad$$\qquad$$\qquad$ & \psfragscanon  \psfrag{1}[bl]{$\bar{\psi}\mathcal{O}\psi$}  \psfrag{2}{$\gamma(P)$} 
\psfrag{3}{nonlocal}
\psfrag{4}{current}\includegraphics[width=4cm]{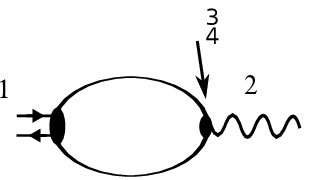}\psfragscanoff\tabularnewline
\end{tabular}
\par\end{centering}

\caption{a) Bag diagram for the radiative $D^{0*}$ vector meson decay. The
lower blob corresponds to photon Distribution Amplitude. b) Simple
quark loop corresponding to photon DA in the quark model. Double
external line represents bilocal quark operator on the light-cone.}

\end{figure}

Photon DAs appear for example in the description of vector mesons
radiative decays. To be more specific consider for instance the process
$D^{0*}\left(q+p\right)\rightarrow D^{0}\left(p\right)+\gamma\left(q\right)$
(Fig. \ref{fig:DA1}a). Using OPE one can then write the amplitude
as products of factors that are divergent on the light-cone and
finite photon-to-vacuum matrix elements. The latter can be identified
with the photon DA as we shall see.

Before we give the more precise definition of the photon DA we should
recall usefull kinematical variables. One defines two null vectors
$n=\left(1,0,0,-1\right)$ and $\tilde{n}=\left(1,0,0,1\right)$.
Then any four-vector $v^{\mu}$ can be decomposed into {}``plus'',
{}``minus'' and transverse components\begin{equation}
v^{\mu}=v^{+}\frac{\tilde{n}^{\mu}}{2}+v^{-}\frac{n^{\mu}}{2}+v_{T}^{\mu}.\label{eq:lightcoord1}\end{equation}

Photon Distribution Amplitude is defined as a Fourier transform of
the photon-to-vacuum matrix element of the non-local quark operator
on the light-cone. In general this can be written as\begin{multline}
\int\frac{d\lambda}{2\pi}\, e^{i(2u-1)\lambda P^{+}}\left\langle 0\left\vert \overline{\psi}\left(\lambda n\right)\mathcal{O}\psi\left(-\lambda n\right)\right\vert \gamma\left(P\right)\right\rangle
\sim F_{\mathcal{O}}
\left(P^{2}\right) \\ \times\left\{ \mathcal{O}_{\mathrm{twist-2}\,}\,\phi_{\mathcal{O}}^{\mathrm{twist-2}}\left(u,P^{2}\right)+\mathcal{O}_{\mathrm{twist-3}}\,\phi_{\mathcal{O}}^{\mathrm{twist-3}}\left(u,P^{2}\right)+\ldots\right\} \label{eq:def_DA}\end{multline}
where $\mathcal{O}=\left\{ \sigma^{\mu\nu},\gamma^{\mu},\gamma^{\mu}\gamma_{5}\right\} $
corresponds to different tensor nature of bilocal operators, $\mathcal{O}_{\mathrm{twist-2}}$,
$\mathcal{O}_{\mathrm{twist-3}}$,~$\ldots$\textit{ }denote apropriate
tensor structures which are multiplied by photon DA $\phi_{\mathcal{O}}$
of given kinematical twist. Notice that we do not assume that the photon
is on-shell. Then the decay constants $F_{\mathcal{O}}$ depend on
photon virtuality $P^{2}$ and become a kind of {}``form factors''
--- we shall use this terminology in the following. For more precise definitions
of the photon DAs refer to \cite{PBall,BronFoton,stareTDA}.

\begin{figure}
\begin{centering}
\label{fig:DA2}\begin{tabular}{ll}
a)  & b)\tabularnewline
\includegraphics[width=6cm]{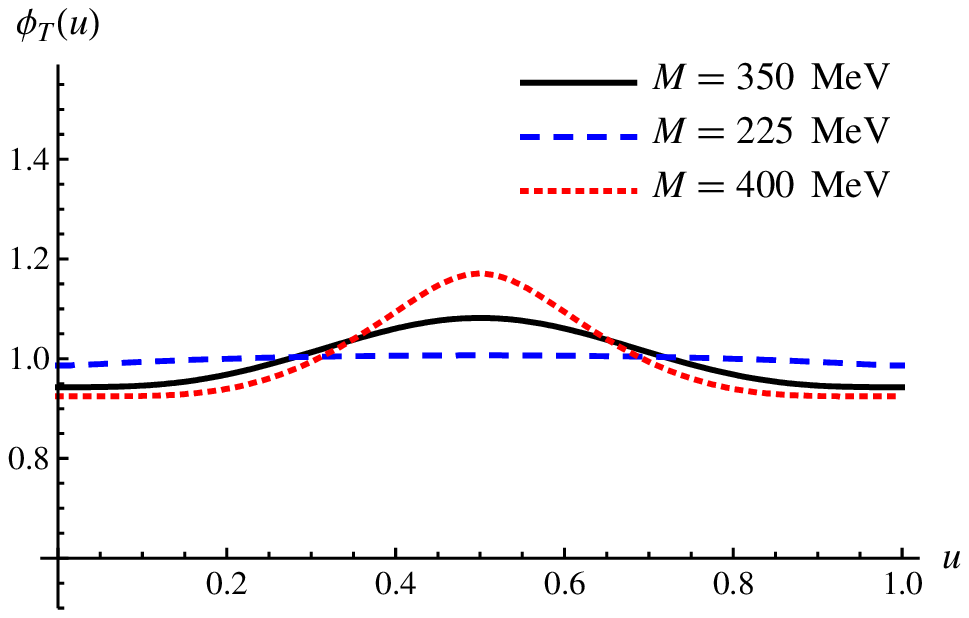} $\,\,$ & \includegraphics[width=6cm]{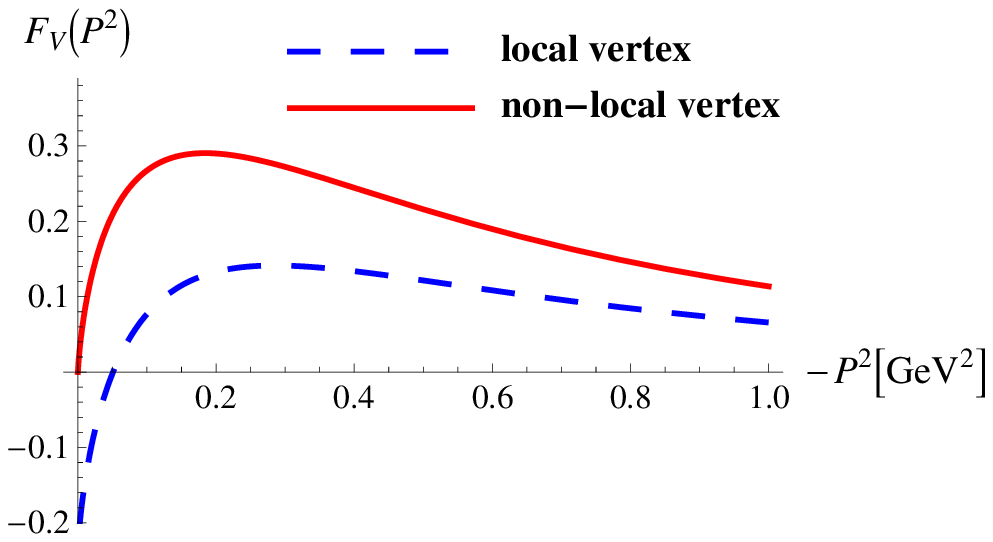}\tabularnewline
\end{tabular}
\par\end{centering}

\caption{a) Twist-2 tensor photon Distribution Amplitude for $n=1$ and several values
of the constituent quark mass $M_{0}$. b) Vector form
factor for $M_{0}=350\,\mathrm{MeV}$,
$n=1$, calculated using naive vector vertex $\gamma^{\mu}$ (dashed
line) and the modified one $\Gamma^{\mu}$ (solid). Notice that the
modified vertex assures that $F_{V}$ vanishes for real photon as
required by QED.}

\end{figure}

Using the non-local chiral quark model described in Section \ref{sec:model}
we calculated photon DAs up to twist-4 in tensor, vector and axial
channels \cite{nasz_foton} in one loop approximation (Fig. \ref{fig:DA1}b).
Our results are analytical up to the solution of a certain polynomial
equation. Let us briefly summarize our results. For real photon the
leading DA is the twist-2 tensor amplitude $\phi_{\sigma^{\mu\nu}}\equiv\phi_{T}$
(Fig. \ref{fig:DA2}a) corresponding to $\sigma^{\mu\nu}$ structure.
We find it is almost flat and non-vanishing in the end-points. Also
the sensitivity to the $n$ parameter, \textit{i.e.} to the shape of $F\left(k\right)$
is rather small. In the vector channel one has to subtract the infinite,
perturbative part when calculating the corresponding matrix element. Then
we find in particular that leading twist vector DA vanishes in the
end-points. It can be easily shown on general grounds that the vector
{}``form factor'' $F_{\gamma^{\mu}}\left(P^{2}\right)\equiv F_{V}\left(P^{2}\right)$
should be zero for the real photon. This property is maitained in
our model only when we use modified vector vertex (Fig. \ref{fig:DA2}b),
as described in Section \ref{sec:model}. Higher twist amplitudes
turn out to be rather strongly model dependend. Moreover some of
them contain Dirac delta functions in the end points, they should
be therefore viewed rather as the generalized functions. Similar calculation
was previously done in Ref. \cite{BronFoton} and differs from ours
in some points.

The left hand side of the definition \eqref{eq:def_DA} is dimensionfull,
therefore we should have several quantities that set up the characteristic
mass scale for photon DAs. Among others, it is a quark condensate,
already disscussed in Section \ref{sec:model}. The non-local chiral
quark model with \eqref{Fkdef} allows to obtain the following {}``analytical''
expression for the quark condensate\begin{equation}
\left\langle \bar{q}q\right\rangle =-\frac{N_{c}M_{0}^{2}\Lambda_{n}^{2}}{4\pi^{2}}\sum_{i=1}^{4n+1}f_{i}\eta_{i}^{2n}\left(1+\eta_{i}\right)\,\ln\left(1+\eta_{i}\right),\label{eq:condensate1}\end{equation}
where the complex numbers $\eta_{i}$ are numerical solutions to the
equation $z^{4n+1}+z^{4n}-\left(M_{0}/\Lambda_{n}\right)^{2}=0$,
while $f_{i}$ are defined as $f_{i}=\prod_{k\neq i}^{4n+1}\left(\eta_{i}-\eta_{k}\right)^{-1}$.
For example for $M_{0}=350\,\mathrm{MeV}$ and $n=1$ we get $\left\langle \bar{q}q\right\rangle =\left(-253\,\mathrm{MeV}\right)^{3}$.
It turns out that in general the values of $\left\langle \bar{q}q\right\rangle $
rather strongly depend on model parameters.

\section{Pion-photon Transition Distribution Amplitudes}

\label{sec:TDA}

In this section we switch to more involved applications of the non-local
chiral quark model. Transition Distribution Amplitudes, apart from being
interesting on their own, can serve as a demanding testing ground for
the model. The reason is that they involve diagrams responsible
for axial anomaly. We shall come back to this point later in this section.

\begin{figure}
\begin{centering}
\begin{tabular}{ll}
a)  & b)\tabularnewline
\psfragscanon \psfrag{5}[cc][cc][0.8]{TDA} \psfrag{6}{$\gamma^*(q_2)$} \psfrag{1}{$\pi^-(q_1)$} \psfrag{2}{$\pi^+(P_1)$} \psfrag{7}{$\gamma(P_2)$} \psfrag{3}{$u$} \psfrag{4}{$\bar{d}$} \psfrag{8}{$e^-$} \psfrag{9}{$e^-$}\includegraphics[width=4.5cm]{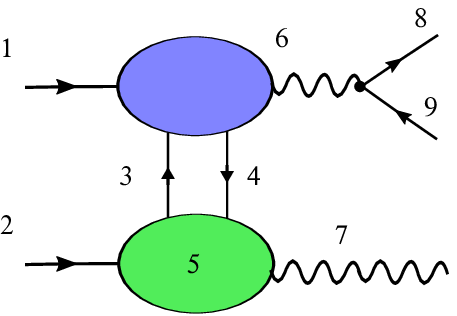}\psfragscanoff
$\qquad$$\qquad$ & \psfragscanon  
\psfrag{3}[cb]{$\bar{\psi}\mathcal{O}\psi$}
\psfrag{1}{$\pi^+(P_1)$}
\psfrag{2}{$\gamma(P_2)$}
\psfrag{4}{nonlocal}
\psfrag{5}{currents}\includegraphics[width=4cm]{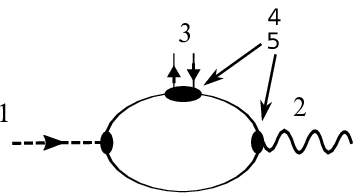}\psfragscanoff\tabularnewline
\end{tabular}
\par\end{centering}

\caption{a) The bag diagram for the process $\pi^{+}\pi^{-}\rightarrow\gamma^{*}\gamma$.
The lower bag represents Transition Distribution Amplitude while the
upper corresponds to the hard process. b) The quark loop corresponding
to TDA, the bilocal operator is assumed to {}``live'' on the light-cone.
In order to recover correct normalization both vertices have to be
non-local.}

\label{fig:TDA1}
\end{figure}

Transition Distribution Amplitudes were originally introduced in order
to describe hadron-antihadron annihilation into two photons, \textit{i.e.}
the process $H\bar{H}\rightarrow\gamma^{*}\gamma$ or backward virtual
Compton scattering $\gamma^{*}H\rightarrow\gamma H$ \cite{Pire}.
The amplitudes for these processes can be described in QCD analogously
to the reactions $H\bar{H}\rightarrow\gamma^{*}$ and $\gamma^{*}H\rightarrow H$
respectively with the restriction that Distribution Amplitudes for
$H$ should be replaced by a new object - Transition Distribution
Amplitudes (Fig. \ref{fig:TDA1}a). First estimates were done in Refs.
\cite{Tiburzi,BronTDA,Courtoy,stareTDA}.

Before we give the general definition of TDAs we should define relevant
kinematics. We consider pion with momentum $P_{1}^{\mu}$ transforming
into the photon with momentum $P_{2}^{\mu}$. We define the momentum
transfer as $q^{\mu}=P_{2}^{\mu}-P_{1}^{\mu}$ and the momentum transfer
squared $t=q^{2}$ which is assumed to be small. Using the average
momentum $p^{\mu}=\frac{1}{2}\left(P_{1}^{\mu}+P_{2}^{\mu}\right)$
we define so called skewedness $\xi=-q^{+}/2p^{+}$, which is a standard
variable in the GPDs formalism. We consider chiral limit and real
photons, i.e. $P_{1}^{2}=P_{2}^{2}=0$. 

The general definition of leading twist TDAs can be written as\begin{multline}
\int\frac{d\lambda}{2\pi}e^{i\lambda Xp^{+}}\left\langle \gamma\left(P_{2},\varepsilon\right)\left\vert \overline{\psi}\left(\lambda n\right)\mathcal{O}\psi\left(-\lambda n\right)\right\vert \pi^{+}\left(P_{1}\right)\right\rangle =\mathcal{O_{\mathrm{twist-2}}}\, D\left(X,\xi,t\right)+\ldots,\label{eq:def_TDA}\end{multline}
where in practice $\mathcal{O}=\left\{ \gamma^{\mu},\gamma^{\mu}\gamma_{5}\right\} $.
Dots stand for the other terms that can appear and are not related
to TDA under consideration. For example in the axial channel, \textit{i.e.}
for $\mathcal{O}=\gamma^{\mu}\gamma_{5}$, pion DA accompanied by
massles pole appears on the right hand side. This reflects the fact
that the axial current couples to a pion directly. In the following
we denote vector TDA as $V\left(X,\xi,t\right)$ (\textit{i.e.} for $\mathcal{O}=\gamma^{\mu}$)
and the axial TDA as $A\left(X,\xi,t\right)$ (for $\mathcal{O}=\gamma^{\mu}\gamma_{5}$).

There is very important property that TDAs should posses, namely
so called polynomiality\begin{equation}
\int_{-1}^{1}dX\, X^{n}D\left(X,\xi,t\right)=a_{n}\left(t\right)\xi^{n}+a_{n-1}\left(t\right)\xi^{n-1}+\ldots+a_{0}\left(t\right),\label{eq:polynomial}\end{equation}
which follows simply from Lorentz invariance. In principle the zeroth
moment is related to the corresponding form factor. Second very important
constraint is the normalization of the vector TDA, which is fixed by the axial anomaly\begin{equation}
\int_{-1}^{1}dX\, V\left(X,\xi,t=0\right)=\frac{1}{2\pi^{2}}.\label{eq:normaliz}\end{equation}
Above condition is model independent and can be derived using Ward-Takahashi
identities that relate the two-photon matrix elements of the axial
and pseudoscalar currents. The latter can be then identified with
our matrix element \eqref{eq:def_TDA} with $\mathcal{O}=\gamma^{\mu}$.
There is no similar normalization condition for axial TDA.
However, in the local models, \textit{i.e.} with $M\left(k\right)\equiv M$
it turns out that \begin{equation}
\int_{-1}^{1}dX\, A_{\mathrm{local}}\left(X,\xi,t=0\right)=\int_{-1}^{1}dX\, V\left(X,\xi,t=0\right)=\frac{1}{2\pi^{2}}.\label{eq:normaliz_axial}\end{equation}

\begin{figure}
\begin{centering}
\label{fig:TDA2}\begin{tabular}{ll}
a)  & b)\tabularnewline
\includegraphics[width=6cm]{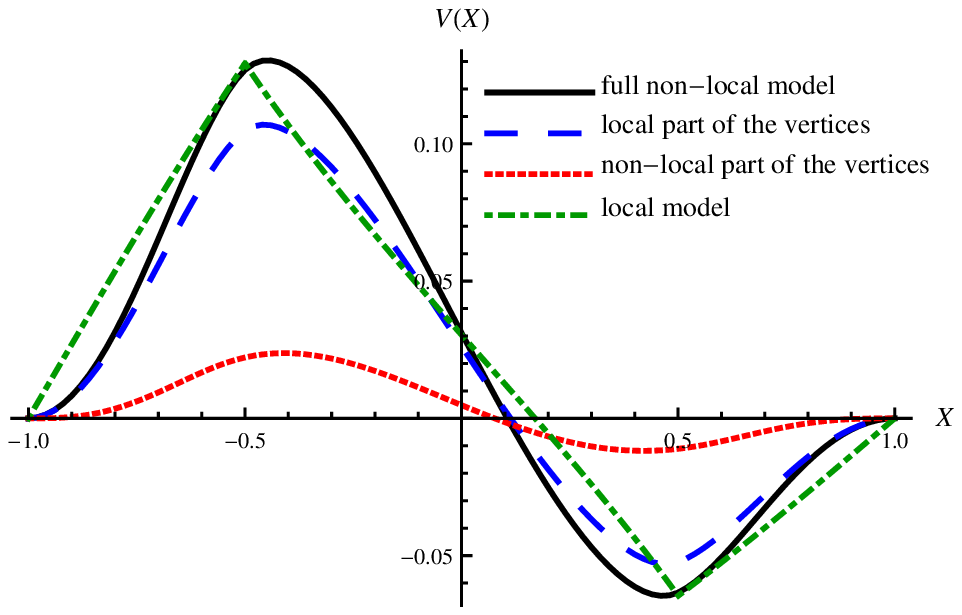} $\,\,$ & \includegraphics[width=6cm]{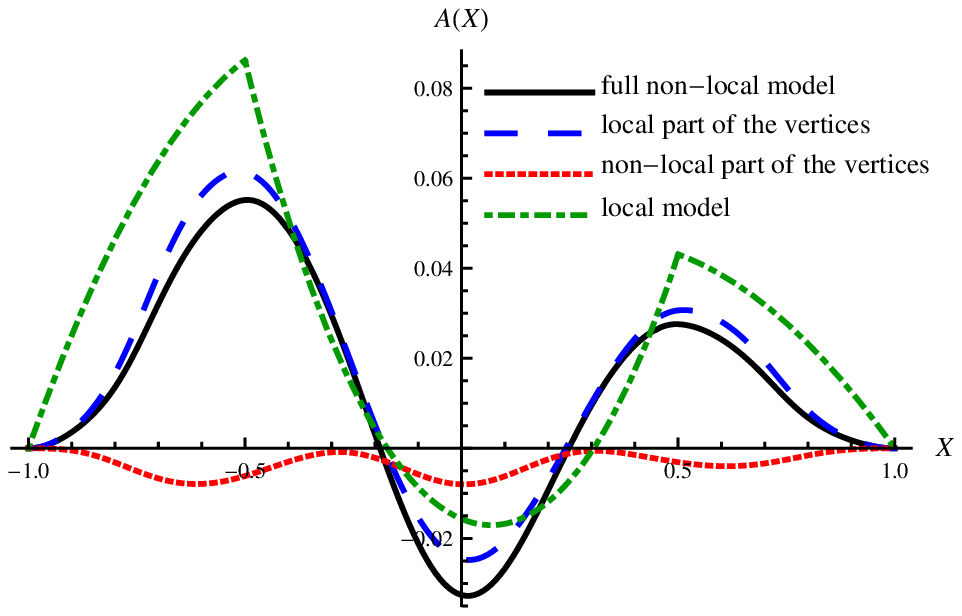}\tabularnewline
\end{tabular}
\par\end{centering}

\caption{a) Vector Transition Distribution Amplitude for $M=350\,\mathrm{MeV}$,
$n=1$, $t=-0.1\,\mathrm{GeV}^{2}$ and $\xi=0.5$. Solid line corresponds
to the full non-local model with non-local vertices and is a sum of the
dashed line and the dotted. Dash-dotted line was obtained in local
model, \textit{i.e.} with $M\left(k\right)\equiv M$. b) The same for the axial TDA.
Here the addition coming from the non-local part of the vertices gives
negative contribution.}

\end{figure}

In the quark model, calculation of the TDAs reduces to performing
quark loop shown in Fig. \ref{fig:TDA1}b. We present a typical
results in Fig. \ref{fig:TDA2} \cite{nowaTDA}. Notice first
that curves obtained in non-local model are much more smooth then
the ones obtained in the local model. Next, we find that the normalization
condition \eqref{eq:normaliz} is recovered only when light-cone bilocal
current in \eqref{eq:def_TDA} is also modified according to \eqref{eq:vector_vertex}.
At the same time the normalization of the axial TDA is much lower
than \eqref{eq:normaliz_axial}. This result is important because
zeroth moments of vector and axial TDAs are directly related to the
vector and axial form factors which can be estimated experimentally.
To be more precise the relation is\begin{equation}
{\displaystyle \int\limits _{-1}^{1}}dX\,\left\{ \begin{array}{c}
V\left(X,\xi,t\right)\\
A\left(X,\xi,t\right)\end{array}\right.=2\sqrt{2}F_{\pi}\left\{ \begin{array}{c}
F_{V}^{\chi}\left(t\right)\\
F_{A}^{\chi}\left(t\right)\end{array}\right.,\label{eq:sumrules}\end{equation}
where the superscript $\chi$ denotes that these quantities are defined
in the chiral limit. The experimental values for $t=0$ are (PDG)\begin{equation}
F_{V}^{\mathrm{exp}}\left(0\right)=0.017\pm0.008,\label{eq:FVexp}\end{equation}
\begin{equation}
F_{A}^{\mathrm{exp}}\left(0\right)=0.0115\pm0.0005,\label{eq:FAexp}\end{equation}
\begin{equation}
\left(F_{A}\left(0\right)/F_{V}\left(0\right)\right)_{\mathrm{exp}}=0.7_{-0.2}^{+0.6}.\label{eq:Fratio}\end{equation}
On the other hand the normalization \eqref{eq:normaliz} gives (model
independent)\begin{equation}
F_{V}^{\chi}\left(0\right)\approx0.027,\label{eq:F3}\end{equation}
what overshoots \eqref{eq:FVexp} more than one standard deviation.
The results for axial form factor are model dependent. For reasonable
model parameters we obtain:
\begin{center}
\begin{tabular}{|c|c|c|c|}
\hline 
$M\,\left[\mathrm{MeV}\right]$  & $n$  & $F^{\chi}_{A}\left(0\right)$  & $F^{\chi}_{A}\left(0\right)/F^{\chi}_{V}\left(0\right)$\tabularnewline
\hline
\hline 
225  & 1  & 0.0217  & 0.80\tabularnewline
\hline 
350  & 1  & 0.0168  & 0.62\tabularnewline
\hline 
350  & 5  & 0.0163  & 0.60\tabularnewline
\hline 
400  & 1  & 0.0161  & 0.60\tabularnewline
\hline 
400  & 5  & 0.0152  & 0.56\tabularnewline
\hline
\end{tabular}
\end{center}
We see that indeed the assumption of non-locality \eqref{eq:const_mass}
lowers the value of $F^{\chi}_{A}$ towards the experimental data.

\begin{figure}
\begin{centering}
\includegraphics[width=7.5cm]{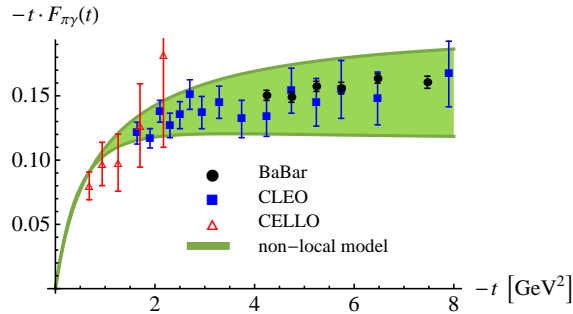}
\par\end{centering}

\caption{The experimental data for pion-photon transition form factor $F_{\pi\gamma}$
times momentum transfer $t$. The shaded area represents the predictions
from the non-local chiral quark model predictions for sensible model parameters.
We get the best description of the low momentum tranfer data for
$M_{0}\sim 300\,\mathrm{MeV}$.}

\label{fig:TDA3}
\end{figure}

Moreover $F_{V}^{\chi}$ is directly related to so called pion-photon
transition form factor $F_{\pi\gamma}$ via the relation\begin{equation}
F_{\pi\gamma}\left(t\right)=\sqrt{2}F_{V}^{\chi}\left(t\right).\label{eq:transff}\end{equation}
This quantity describes the pion decay $\pi^{0}\rightarrow\gamma^{\ast}\gamma$
process and was measured by CLEO \cite{cleo}, CELLO \cite{cello}
and recently by BaBar \cite{babar} collaborations. We compare our
predictions to the experimental ones in Fig. \ref{fig:TDA3}. There
is however important remark in order. Notice, that by definition TDAs
are sensible only for small momentum transfers $t$, the precise range
of application is however not known. Therefore, as an example we have chosen
arbitrarliy the range of $0-8\,\mathrm{GeV^{2}}$. It is worth noting at this point
that the new BaBar data are in disagreement with the standard QCD factorization formula, as it was already
remarked in Introduction. In QCD, the pion-photon form factor can be described using pion DA and some perturbatively 
calculable factor, which leads to the certain asymptotic form. New BaBar data cover the range of $0-40\, \mathrm{GeV}^2$ (in Fig. \ref{fig:TDA3} we 
retained only the relevant low momentum data) and cross the asymptotic line already at about $10\, \mathrm{GeV}^2$.
One way to resolve this discrepancy is to note that pion DA which vanishes at the end-points was assumed in the standard factorization formula. In Refs. \cite{RadyushkinBABAR,PolyakovBABAR} the autors study the flat pion DA in order to describe the new BaBar data (but see also \cite{Stefanis}).

\section{Summary}

Let us briefly summarize our presentation. In the begining we recalled
chiral quark models, starting from widely known Nambu---Jona-Lasinio
model. We argued that spontaneous chiral symmetry breaking is their
main ingredient. We also showed that they lead to nonzero
quark condensates and in turn to dynamically generated
constituent quark mass, which in general can depend on momentum. Next
we used a simple ansatz for this dependence and applied the model to
two low-energy objects: photon Distribution Amplitude and photon-pion
Transition Distribution Amplitude. We find that they fulfil most symmetries
required by QCD, provided we modify the vector and axial vertices
in such a way that relevant currents are conserved. We find also that
form factors which are calculated using Transition Distribution Amplitudes
are realistic when compared to the experimental data.

At the end we draw attention to important issues which was not covered by this
presentation. First of all, in QCD all the low-energy quantities depend on some
factorization scale $\mu$ and are a subject for corresponding QCD evolution.
On the other hand, within effective models they are obtained at some fixed $\mu$, which is
in fact unknown (although can be roughly estimated). Therefore before one makes a real use
of them the evolution has to be applied. The second remark is rather a technical one and concerns
the cutoff $\Lambda_n$ parameter in \eqref{Fkdef}. One should not confuse it with the
scale $\mu$ of the model, as discussed in \cite{Rostw}.

\section*{Acknowledgements}
Based on work done in collaboration with Micha{\l} Prasza{\l}owicz. The Author
acknowledges support the Polish-German cooperation agreement between 
Polish Academy of Sciences and DFG.

\end{document}